\documentclass[final]{siamltex}

\usepackage[dvips]{graphicx}
\usepackage{epsf,ams}

\newtheorem{thm}{Theorem}
\newtheorem{conj}[thm]{Conjecture}

\newcommand {\ve}{{\rm{vec}}}
\renewcommand{\R}{\mathbb R}

\title{A measure of similarity between graph vertices: Applications
to synonym extraction and web searching\thanks{This paper presents
results that have been obtained in part during research performed
by AG, MH and PS under the guidance of VB and PVD. This research
was supported by the National Science Foundation under Grant
No.~CCR 99-12415, by the University of Louvain under a grant
FSR2000, and by the Belgian Programme on Inter-university Poles of
Attraction, initiated by the Belgian State, Prime Minister's
Office for Science, Technology and Culture. The scientific
responsibility rests with its authors.}}

\author{Vincent D. Blondel\thanks{Division of Applied Mathematics,
Universit\'e catholique de Louvain, 4 Ave. G. Lemaitre, B-1348
Louvain-la-Neuve, Belgium ({\tt blondel@inma.ucl.ac.be}), {\tt
http://www.inma.ucl.ac.be/$\sim$blondel}.} \and Anah\'i
Gajardo\thanks{Departamento de Ingeniería Matem\'atica,
Universidad de Concepci\'on, Casilla 160-C,Concepci\'on, Chile
({\tt anahi@ing-mat.udec.cl}).} \and Maureen Heymans\thanks{Google
Inc., 2400 Bayshore Parkway, Mountain View, CA  94043, USA ({\tt
maureen@google.com}).}     \and Pierre Senellart\thanks{Computer
Science Department, \'Ecole normale sup\'erieure, 45 rue d'Ulm,
F-75230 Paris cedex 05, France (\tt Pierre.Senellart@ens.fr).}
        \and
        Paul Van Dooren\thanks{Division of Applied Mathematics, Universit\'e
catholique de Louvain, 4 Ave. G. Lemaitre, B-1348
Louvain-la-Neuve, Belgium ({\tt vdooren@csam.ucl.ac.be}), {\tt
http://www.inma.ucl.ac.be/$\sim$vdooren}.} }

\begin{document}

\maketitle

\begin{abstract}
We introduce a concept of {similarity} between vertices of
directed graphs. Let $G_A$ and $G_B$ be two directed graphs with
respectively $n_A$ and $n_B$ vertices. We define a $n_B \times
n_A$ {\it similarity matrix} $\mathbf{S}$ whose real entry
$s_{ij}$ expresses how similar vertex $j$ (in $G_A)$ is to vertex
$i$ (in $G_B$)~: we say that $s_{ij}$ is their {\it similarity
score}. The similarity matrix can be obtained as the limit of the
normalized even iterates of $S(k+1)=BS(k)A^T+B^TS(k)A$ where $A$
and $B$ are adjacency matrices of the graphs and $S(0)$ is a
matrix whose entries are all equal to one. In the special case
where $G_A=G_B=G$, the matrix $\mathbf{S}$ is square and the score
$s_{ij}$ is the similarity score between the vertices $i$ and $j$
of $G$. We point out that Kleinberg's ``hub and authority" method
to identify web-pages relevant to a given query  can be viewed as
a special case of our definition in the case where one of the
graphs has two vertices and a unique directed edge between them.
In analogy to Kleinberg, we show that our similarity scores are
given by the components of a dominant eigenvector of a
non-negative matrix. Potential applications of our similarity
concept are numerous. We illustrate an application for the
automatic extraction of synonyms in a monolingual dictionary.
\end{abstract}



\pagestyle{myheadings} \thispagestyle{plain} \markboth{V. BLONDEL,
A. GAJARDO, M. HEYMANS, P. SENELLART,  AND P. VAN
DOOREN}{SIMILARITY IN GRAPHS}


\section{Generalizing hubs and authorities}
\label{s1}

Efficient web search engines such as Google are often based on the
idea of characterizing the {most important} vertices in a graph
representing the connections or links between pages on the web.
One such method,  proposed by Kleinberg \cite{klein}, identifies
in a set of pages relevant to a query search the subset of pages
that are good \emph{hubs} or the subset of pages that are good
\emph{authorities}. For example, for the query ``university", the
home-pages of Oxford, Harvard and other universities are good
authorities, whereas web pages that point to these home-pages are
good hubs.  Good hubs are pages that point to good authorities,
and good authorities are pages that are pointed to by good hubs.
From these implicit relations, Kleinberg derives an iterative
method that assigns an ``authority score'' and a ``hub score'' to
every vertex of a given graph. These scores can be obtained as the
limit of a converging iterative process which we now describe.

Let $G=(V,E)$ be a graph with vertex set $V$ and with edge set $E$
and let $h_j$ and $a_j$ be the hub and authority scores of vertex
$j$. We let these scores be initialized by some positive values
and then update them simultaneously for all vertices according to
the following \emph{mutually reinforcing relation}~: the hub score
of vertex $j$ is set equal to the sum of the authority scores of
all vertices pointed to by $j$ and, similarly, the authority score
of vertex $j$ is set equal to the sum of the hub scores of all
vertices pointing to $j$~:
$$
\left\{
\begin{array}{rclll}
h_{j} & \leftarrow & \sum_{i: (j,i) \in E} a_{i}\\[2mm]
a_{j} & \leftarrow &  \sum_{i: (i,j)\in E} h_{i}
\end{array}
\right.
$$

Let $B$ be the matrix  whose entry $(i, j)$ is equal to the number
of edges between the vertices $i$ and $j$ in $G$ (the
\emph{adjacency matrix} of $G$), and let $h$ and $a$ be the
vectors of hub and authority scores. The above updating equations
then take the simple form
$$
\left[ \begin{array}{c} h \\ a \end{array} \right]_{k+1} = \left[
\begin{array}{cc} 0 & B \\ B^T & 0 \end{array} \right] \left[
\begin{array}{c} h \\ a \end{array} \right]_{k}, \qquad k=0, 1,
\ldots$$ which we denote in compact form by
$$x_{k+1}=M\; x_k, \qquad k=0, 1, \ldots$$ where
$$
\label{e11} x_k = \left[ \begin{array}{c} h \\ a \end{array}
\right]_{k}, \quad M = \left[ \begin{array}{cc} 0 & B \\ B^T & 0
\end{array} \right].
$$
Notice that the matrix $M$ is symmetric and
non-negative\footnote{A  matrix or a vector $Z$ will be said
 non-negative (positive) if all its components are
non-negative (positive), we write $Z\geq 0$ ($Z >0$) to denote
this.}. We are only interested in the relative scores and we will
therefore consider the \emph{normalized} vector sequence
\begin{equation}
\label{a1}
 z_0=x_0>0, \quad z_{k+1}=\frac{M z_k}{\|M z_k\|_2}, \quad
k=0, 1, \ldots
\end{equation}
where $\|\cdot \|_2$ is the Euclidean vector norm. Ideally, we
would like to take the limit of the sequence $z_k$ as a definition
for the hub and authority scores. There are two difficulties with
such a definition.

A first difficulty is that the sequence $z_k$ does not always
converge. In fact, sequences associated with non-negative matrices
$M$ with the above block structure almost never converge but
rather oscillate between the limits
$$
z_{even}  =  \lim_{k\rightarrow \infty} z_{2k} \quad \mbox{ and }
\quad z_{odd}  =  \lim_{k\rightarrow \infty} z_{2k+1}.
$$
 We prove in Theorem \ref{tt2} that this is true in general for
symmetric non-negative matrices, and that either the sequence
resulting from (\ref{a1}) converges, or it doesn't and then the
even and odd sub-sequences do converge. Let us consider both
limits for the moment.

The second difficulty is that the limit vectors $z_{even}$ and
$z_{odd}$ do in general depend on the initial vector $z_0$ and
there is no apparently natural choice for $z_0$. The set of all
limit vectors obtained when starting from a positive initial
vector is given by
$$Z=\{z_{even}(z_0), z_{odd}(z_0): z_0 >0 \}$$
and we would like to select one particular vector in that set. The
vector $z_{even}$ obtained for $z_0={\bf 1}$ (we denote by ${\bf
1}$ the vector, or matrix,  whose entries are all equal to 1) has
several interesting features that qualifies it as a good choice:
it is particularly easy to compute, it possesses several nice
properties (see in particular Section \ref{s5}), and it has the
extremal property, proved in Theorem \ref{t2}, of being the unique
vector in $Z$ of largest possible 1-norm (the 1-norm of a vector
is the sum of all the magnitudes of its entries). Because of these
features, we take the two sub-vectors of $z_{even}({\bf 1})$ as
definitions for the hub and authority scores. In the case of the
above matrix $M$, we have
$$
M^2 = \left[ \begin{array}{cc} BB^T & 0\\ 0 & B^TB
\end{array}\right]
$$
and from this equality it follows that, if the dominant invariant
subspaces associated with $BB^T$ and $B^TB$ have dimension one,
then the normalized hub and authority scores are simply given by
the normalized dominant eigenvectors of $BB^T$ and $B^TB$. This is
the definition used in \cite{klein} for the authority and hub
scores of the vertices of $G$. The arbitrary choice of $z_0={\bf
1}$ made in \cite{klein} is shown here to have an extremal norm
justification. Notice that when the invariant subspace has
dimension one, then there is nothing particular about the starting
vector ${\bf 1}$ since any other positive vector $z_0$ would give
the same result.

\begin{figure}
\begin{center}
\begin{tabular}{cc}
\begin{minipage}{4.5cm}\includegraphics[scale=0.4]{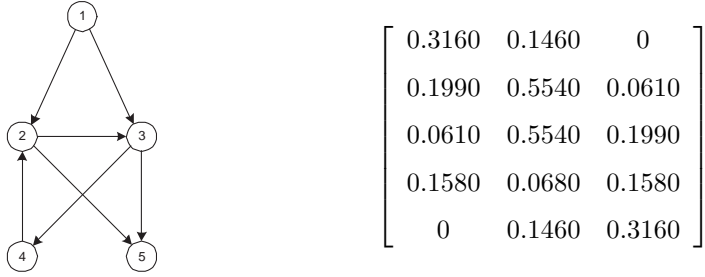}\end{minipage} &
$\left[\begin {array}{ccc}
  0.3160& 0.1460&0\\\noalign{\medskip} 0.1990& 0.5540&
0.0610\\\noalign{\medskip} 0.0610& 0.5540&
0.1990\\\noalign{\medskip} 0.1580& 0.0680&
0.1580\\\noalign{\medskip}0& 0.1460& 0.3160\end {array}\right]$
\end{tabular}
\end{center}

\label{fig:smm} \caption{A graph and its similarity matrix with
the structure graph $1\longrightarrow 2\longrightarrow 3$. The
similarity score of vertex 3 with vertex 2 of the structure graph
is equal to 0.5540. }
\end{figure}

We now generalize this construction. The authority score of vertex
$j$ of $G$ can be thought of as a similarity score between vertex
$j$ of $G$ and vertex \emph{authority} of the graph
$$hub \longrightarrow authority$$
and, similarly, the hub score of vertex $j$ of $G$ can be seen as
a similarity score between vertex $j$ and vertex \emph{hub}. The
mutually reinforcing updating iteration used above can be
generalized to graphs that are different from the  hub-authority
structure graph. The idea of this generalization is easier to
grasp with an example; we illustrate it first on the path graph
with three vertices and then provide a definition for arbitrary
graphs. Let $G$ be a graph with edge set $E$ and  adjacency matrix
$B$ and consider the \emph{structure graph}
$$1\longrightarrow 2\longrightarrow 3.$$
With each vertex $j$ of $G$ we now associate three scores $x_{i1},
x_{i2}$ and $x_{i3}$; one for each vertex of the structure graph.
We initialize these scores with some positive value and then
update them according to the following mutually reinforcing
relation
$$
\left\{ \begin{array}{rclll}
x_{i1} & \leftarrow & & \sum_{j: (i, j) \in E} x_{i2}\\[2mm]
x_{i2} & \leftarrow &  \sum_{j: (j, i)\in E} x_{i1} & & +\sum_{j:
(i, j)\in E} x_{i3}\\[2mm]
x_{i3} & \leftarrow &  & \sum_{j: (j, i)\in E}  x_{i2}
\end{array} \right.
$$
or, in matrix form (we denote by $\mathbf{x}_j$ the column vector
with entries $x_{ij}$),
$$\left[
\begin{array}{c}
\mathbf{x}_1 \\ \mathbf{x}_2 \\ \mathbf{x}_3
\end{array}
\right]_{k+1} = \left[
\begin{array}{ccc}
0 & B & 0\\
B^T & 0 & B\\
0 & B^T & 0
\end{array}
\right] \left[
\begin{array}{c}
\mathbf{x}_1 \\ \mathbf{x}_2 \\ \mathbf{x}_3
\end{array}
\right]_{k}, \qquad k=0,1,\ldots
$$
which we again denote by $x_{k+1}= Mx_k$. The situation is now
identical to that of the previous example and all convergence
arguments given there apply here as well. The matrix $M$ is
symmetric and non-negative, the normalized even and odd iterates
converge and the  limit $z_{even}({\bf 1})$ is among  all possible
limits the unique vector with largest possible 1-norm. We take the
three components of this extremal limit $z_{even}({\bf 1})$  as
definition for the similarity scores $s_1, s_2$ and $s_3$ and
define the similarity matrix by ${\bf S}=[ s_1 \; s_2 \; s_3 ]$. A
numerical example of such a similarity matrix is shown in Figure
\ref{fig:smm}. Note that we shall prove in Theorem \ref{tt6} that
the score $s_2$ can be obtained more directly from $B$ by
computing the dominating eigenvector of the matrix $BB^T+B^TB$.

We now come  to a description of the general case. Assume that we
have two directed graphs $G_A$ and $G_B$ with $n_A$ and $n_B$
vertices and edge sets $E_A$ and $E_B$. We think of $G_A$ as a
structure graph that plays the role of the graphs $hub
\longrightarrow authority$ and $1\longrightarrow 2\longrightarrow
3$ in the above examples.  We consider real scores $x_{i j}$ for
$i=1, \ldots, n_B$ and $j=1, \ldots, n_A$ and simultaneously
update all scores according to the following updating equations

$$
\label{r4}
 x_{ij} \; \leftarrow \sum_{r: (r, i) \in E_B, \; s: (s, j) \in E_A
}  x_{rs}  + \sum_{r: (i, r) \in E_B , \; s: (j, s) \in E_A}
x_{rs}.
$$

This equation can be given an interpretation in terms of the
product graph of $G_A$ and $G_B$. The {\emph product graph} of
$G_A$ and $G_B$ is a graph that has $n_A . n_B$ vertices and that
has an edge between vertices $(i_1, j_1)$ and $(i_2, j_2)$ if
there is an edge between $i_1$ and $i_2$ in $G_A$ and there is an
edge between $j_1$ and $j_2$ in $G_B$. The updating equation
(\ref{a1}) is then equivalent to replacing the values of all
vertices of the product graph by the values of the outgoing and
incoming vertices in the graph.

\begin{figure}
{\small
\begin{center}
\begin{tabular}{cc}
\begin{minipage}{6cm}\includegraphics[scale=0.4]{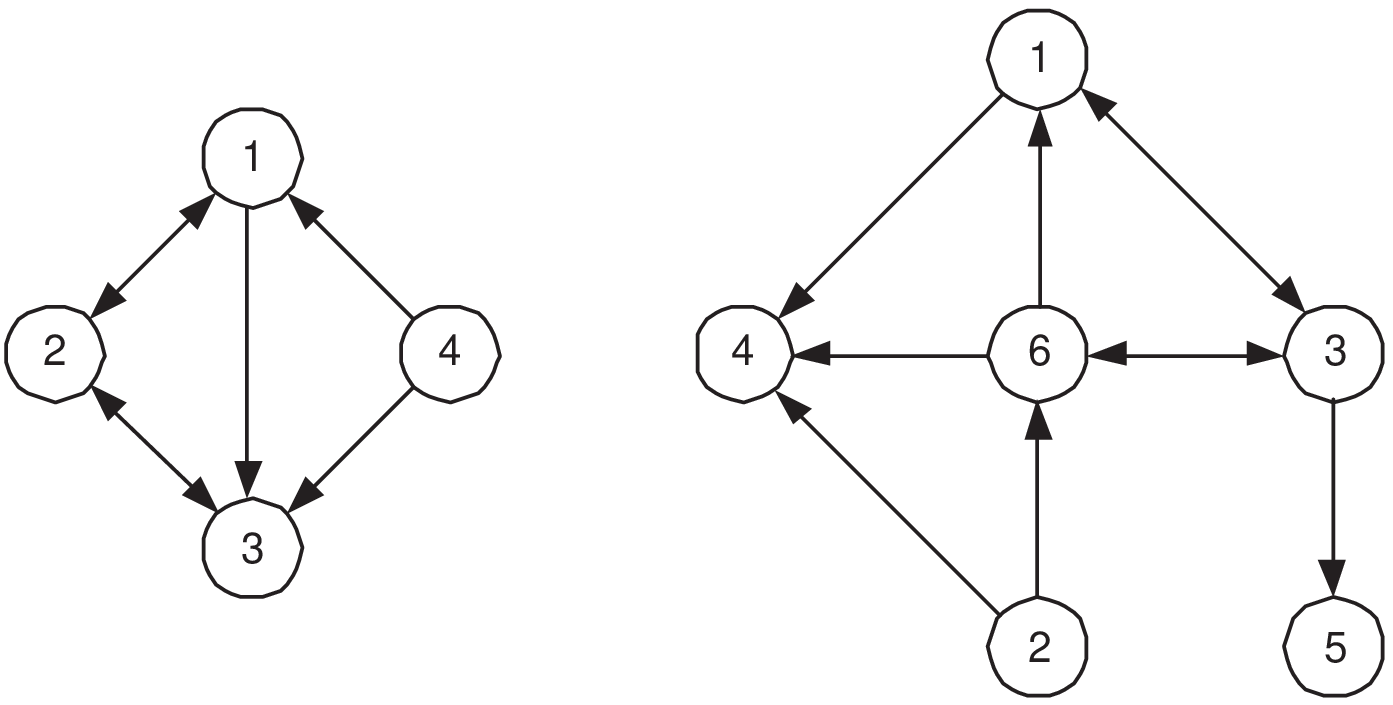}\end{minipage} &
$\left[\begin {array}{cccc}  0.2720& 0.2840& 0.2700&
0.1440\\\noalign{\medskip} 0.1400& 0.1390& 0.0670&
0.1390\\\noalign{\medskip} 0.2700& 0.3120& 0.2560&
0.1650\\\noalign{\medskip} 0.2300& 0.2440& 0.2870&
0.0720\\\noalign{\medskip} 0.0660& 0.0720&
0.1030&0\\\noalign{\medskip} 0.2540& 0.2480& 0.2340& 0.1460
\end {array}\right]$
\end{tabular}
\end{center}
}

\caption{Two graphs $G_A, G_B$ and their similarity matrix. The
vertex of $G_A$ which is most similar to vertex 5 in $G_B$ is
vertex $3$.} \label{f12}
\end{figure}

Equation (\ref{a1}) can also be written in more compact matrix
form. Let $X_k$ be the $n_B \times n_A$ matrix of entries $x_{i
j}$ at iteration $k$. Then the updating equations take the simple
form
\begin{equation} \label{a2}
 X_{k+1}=BX_kA^T + B^TX_kA, \qquad k=0, 1, \ldots
\end{equation}
where $A$ and $B$ are the adjacency matrices of $G_A$ and $G_B$.
We prove in Section \ref{s3} that, as for the above examples, the
normalized even and odd iterates of this updating equation
converge, and that the limit $Z_{even}({\bf 1})$ is among all
possible limits the only one with largest 1-norm. We take this
limit as definition of the similarity matrix. An example of two
graphs and their similarity matrix is shown in Figure
\ref{f12}.\\

It is interesting to note that in the database literature similar
ideas have been proposed \cite{flooding}, \cite{simrank}. The
applications there are information retrieval in large databases
with which a particular graph structure can be associated. The
ideas presented in these conference papers are obviously linked to
those of the present paper, but in order to guarantee convergence
of the proposed iteration to a unique fixed point, the iteration
has to be slightly modified using e.g. certain weighting
coefficients. Therefore, the hub and authority score of Kleinberg
is not a special case of their definition. We thank the authors of
\cite{simrank} for drawing our attention to those references.

The rest of this paper is organized as follows. In Section
\ref{s2}, we describe some standard Perron-Frobenius results for
non-negative matrices that are useful in the rest of the paper. In
Section \ref{s3}, we give a precise definition of the similarity
matrix together with different alternative definitions. The
definition immediately translates into an approximation algorithm
and we discuss in that section some complexity aspects of the
algorithm. In Section \ref{s5} we describe similarity matrices for
the situation where one of the two graphs is a path graph of
length 2 or 3. In Section \ref{s7} we consider the special case
$G_A=G_B=G$ for which the score $s_{ij}$ is the similarity between
the vertices $i$ and $j$ in a single graph $G$. Section \ref{s6}
deals with graphs whose similarity matrix have rank one. We prove
there that if one of the graphs is regular or if one of the graphs
is undirected, then the similarity matrix has rank one. Regular
graphs are graphs whose vertices have the same in-degrees, and the
same out-degrees; cycle graphs, for example, are regular.
 In a final section we report results obtained
for the automatic synonym extraction in a dictionary by using the
central score of a graph. A short version of this paper appears as
a conference contribution in \cite{blond2}.

\section{Graphs and non-negative matrices}
\label{s2}

With any directed graph $G=(V,E)$ one can an associate a
non-negative matrix via an indexation of its vertices. The
so-called {\em adjacency matrix} of $G$ is the matrix $B \in
{\mathbb N}^{n\times n}$ whose entry $b_{ij}$ equals the number of
edges from vertex $i$ to vertex $j$. Let $B$ be the adjacency
matrix of some graph $G$; the entry $(B^k)_{ij}$ is equal to the
number of paths of length $k$ from vertex $i$ to vertex $j$. From
this it follows that a graph is strongly connected if and only if
for every pair of indices $i$ and $j$ there is an integer $k$ such
that $(B^k)_{ij}>0$. Matrices that satisfy this property are said
to be {\emph{ irreducible}}.

In the sequel, we shall need the notion of orthogonal projection
on vector subspaces. Let $\mathcal{V}$ be a linear subspace of
$\R^n$ and let $v \in \R^n$. The {\em orthogonal projection} of
$v$ on $\mathcal{V}$ is the vector in $\mathcal{V}$ with smallest
Euclidean distance to $v$. A matrix representation of this
projection can be obtained as follows. Let $\{v_1, \ldots, v_m\}$
be an orthonormal  basis for $\mathcal{V}$ and arrange the column
vectors $v_i$ in a matrix $V$. The projection of $v$ on
$\mathcal{V}$ is then given by $\Pi v= V V^T v$ and the matrix
$\Pi = V V^T$ is the {\em orthogonal projector} on $\mathcal{V}$.
Projectors have the property that $\Pi^2=\Pi$.

The Perron-Frobenius theory \cite{HJ} establishes interesting
properties about the eigenvectors and eigenvalues for non-negative
matrices. Let the largest magnitude of the eigenvalues of the
matrix $M$ (the \emph{spectral radius} of $M$) be denoted by
$\rho(M)$. According to the Perron-Frobenius Theorem, the spectral
radius of a non-negative matrix $M$ is an eigenvalue of $M$
(called the \emph{Perron root}), and there exists an associated
non-negative vector $x \geq 0$ ($x \neq 0$) such that $Mx=\rho x$
(called the \emph{Perron vector}). In the case of symmetric
matrices, more specific results can be obtained.\\

\begin{thm}\label{teo:P-F} Let $M$ be a symmetric non-negative
matrix of spectral radius $\rho$. Then the algebraic and geometric
multiplicity of the Perron root $\rho$ are equal; there is a
non-negative basis $X\ge 0$ for the invariant subspace associated
with the Perron root; and the elements of the orthogonal projector
$\Pi$ on the vector space associated with the Perron root of $M$
are all non-negative.
\end{thm}

\begin{proof} We use the facts that any symmetric non-negative matrix
$M$ can be permuted to a block-diagonal matrix with irreducible
blocks $M_i$ on the diagonal \cite{HJ,BP}, and that the algebraic
multiplicity of the Perron root of an irreducible non-negative
matrix is equal to one. From these combined facts it follows that
the algebraic and geometric multiplicities of the Perron root
$\rho$ of $M$ are equal. Moreover, the corresponding invariant
subspace of $M$ is obtained from the normalized Perron vectors of
the $M_i$ blocks, appropriately padded with zeros. The basis $X$
one obtains
that way is then non-negative and orthonormal. \end{proof} \\

The next theorem will be used to  justify our definition of
similarity matrix between two graphs. The result describes the
limit vectors of sequences associated with symmetric non-negative
linear transformations.\\

\begin{thm}
\label{t2} \label{tt2} Let $M$ be a symmetric non-negative matrix
of spectral radius $\rho$. Let $z_0>0$ and consider the sequence
$$z_{k+1}= M z_k/\|M z_k\|_2, \quad k=0, \ldots$$ Two convergence cases can
occur depending on whether or not $-\rho$ is an eigenvalue of $M$.
When $-\rho$ is not an eigenvalue of $M$, then the sequence $z_k$
simply converges to ${\Pi z_0}/{\| \Pi z_0\|_2}$, where $\Pi$ is
the orthogonal projector on the invariant subspace associated with
the Perron root $\rho$. When $-\rho$ is an eigenvalue of $M$, then
the subsequences $z_{2k}$ and $z_{2k+1}$ converge to the limits
$$z_{even}(z_0)= \lim_{k \rightarrow \infty} z_{2k} =
\frac{\Pi z_0}{\| \Pi z_0 \|_2} \quad  \mbox{ and } \quad
z_{odd}(z_0)= \lim_{k \rightarrow \infty} z_{2k+1} = \frac{\Pi M
z_0}{\| \Pi M z_0 \|_2},$$ where $\Pi$ is the orthogonal projector
on the sums of the invariant subspaces associated with $\rho$ and
$-\rho$. In both cases the set of all possible limits is given by
$$Z= \{z_{even}(z_0), z_{odd}(z_0) : z_0>0 \} =
\{ \Pi z / \|\Pi z \|_2:  z>0 \} $$ and the vector $z_{even}({\bf
1})$ is the unique vector of largest possible 1-norm in that set.
\end{thm}

\begin{proof} We prove only the case where $-\rho$ is an
eigenvalue; the other case is a trivial modification. Let us
denote the invariant subspaces of $M$ corresponding to $\rho$, to
$-\rho$ and to the rest of the spectrum, respectively by
$\mathcal{V}_\rho$, $\mathcal{V}_{-\rho}$ and $\mathcal{V}_\mu$.
Assume that these spaces are non-trivial, and that we have
orthonormal bases for them~:
$$
MV_{\rho}= \rho V_{\rho}, \quad MV_{-\rho}= -\rho V_{-\rho}, \quad
MV_{\mu}= V_{\mu}M_{\mu},
$$
where $M_{\mu}$ is a square matrix (diagonal if $V_\mu$ is the
basis of eigenvectors) with spectral radius $\mu$ strictly less
than $\rho$. The eigenvalue decomposition can then be rewritten in
block diagonal form~:
$$\begin{array}{rcl} M & = & \left[ \begin{array}{ccc} V_{\rho} &
V_{-\rho} & V_{\mu}
\end{array} \right] \left[ \begin{array}{ccc} \rho I & & \\
& -\rho I& \\ & & M_{\mu}\end{array} \right] \left[
\begin{array}{ccc} V_{\rho} & V_{-\rho} & V_{\mu} \end{array}
\right]^T \\
& = & \rho V_{\rho}V_{\rho}^T - \rho V_{-\rho}V_{-\rho}^T +
V_{\mu} M_{\mu}V_{\mu}^T .
\end{array}
$$
It then follows that
$$
  M^{2}  =  \rho^{2} \Pi + V_{\mu} M_{\mu}^{2} V_{\mu}^T
$$
where $\Pi := V_{\rho}V_{\rho}^T + V_{-\rho}V_{-\rho}^T $ is the
orthogonal projector onto the invariant subspace $\mathcal{V}_\rho
\oplus \mathcal{V}_{-\rho}$ of $M^2$ corresponding to $\rho^2$. We
also have
$$
  M^{2k}  =  \rho^{2k} \Pi + V_{\mu} M_{\mu}^{2k} V_{\mu}^T
$$
and since $\rho(M_{\mu})=\mu<\rho$, it follows from multiplying
this by $z_0$ and $Mz_0$ that
$$z_{2k} = \frac{\Pi z_0} {\| \Pi z_0 \|_2} + O(\mu/\rho)^{2k}
$$
and
$$z_{2k+1} = \frac{\Pi Mz_0} {\| \Pi Mz_0 \|_2} + O(\mu/\rho)^{2k}
$$
provided the initial vectors $z_0$ and $Mz_0$ have a non-zero
component in the relevant subspaces, i.e. provided $\Pi z_0$ and
$\Pi M z_0$ are non-zero. But the Euclidean norm of these vectors
equal $z_0^T\Pi z_0$ and $z_0^TM\Pi Mz_0$ since $\Pi^2=\Pi$. These
norms are both non-zero since $z_0>0$ and both $\Pi$ and $M\Pi M$
are non-negative and non-zero.

It follows from the non-negativity of $M$ and the formula for
$z_{even}(z_0)$ and $z_{odd}(z_0)$ that both limits lie in $\{ \Pi
z / \|\Pi z \|_2:  z>0 \}$. Let us now show that every element
$\hat z_0 \in \{ \Pi z / \|\Pi z \|_2:  z>0 \}$ can be obtained as
$z_{even}(z_0)$ for some $z_0>0$.  Since the entries of $\Pi$ are
non-negative, so are those of $\hat z_0$. This vector may however
have some of its entries equal to zero. From $\hat z_0$ we
construct $z_0$ by adding $\epsilon$ to all the zero entries of
$\hat z_0$. The vector $z_0-\hat z_0$ is clearly orthogonal to
$\mathcal{V}_\rho \oplus \mathcal{V}_{-\rho}$ and will therefore
vanish in the iteration of $M^2$. Thus we have $z_{even}(z_0)=\hat
z_0$ for $z_0 >0$, as requested.

We now prove the last statement. The matrix $\Pi$ and all vectors
are non-negative and $\Pi^2=\Pi$, and so,

\[ \left| \left|\frac{\Pi{\mathbf 1}}{\|\Pi{\mathbf 1}\|_2} \right| \right|_1
=\sqrt{{\mathbf 1}^T \Pi^2 {\mathbf 1}}\] and also
\[ \left| \left|\frac{\Pi{z_0}}{\|\Pi{z_0}\|_2}\right| \right|_1
=\frac{{\mathbf 1}^T \Pi^2 {z_0}}{\sqrt{{z_0}^T \Pi^2 {z_0}}} . \]
Applying the Schwarz inequality to $\Pi z_0$ and $\Pi{\mathbf 1}$
yields
\[ |{\mathbf 1}^T \Pi^2 z_0| \leq
\sqrt{z_0^T \Pi^2 z_0} . \sqrt{{\mathbf 1}^T \Pi^2 {\mathbf 1}}.
\]
with equality only when $\Pi z_0 =\lambda \Pi{\mathbf 1}$ for some
$\lambda \in \C$. But since $\Pi z_0$ and $\Pi{\mathbf 1}$ are
both real non-negative, the proof easily follows.
\end{proof}

\section{Similarity  between vertices in graphs}
\label{s3}

We now come to a formal definition of the similarity matrix of two
directed graphs $G_A$ and $G_B$. The updating equation for the
similarity matrix is motivated in the introduction and is given by
the linear mapping:
\begin{equation}
\label{e7}
 X_{k+1}=  BX_kA^T + B^TX_kA, \qquad k=0, 1, \ldots
\end{equation}
where $A$ and $B$ are the adjacency matrices of $G_A$ and $G_B$.
In this updating equation, the entries of $X_{k+1}$ depend
linearly on those of $X_k$. We can make this dependance more
explicit by using the matrix-to-vector operator that develops a
matrix into a vector by taking its columns one by one. This
operator, denoted $\ve$, satisfies the  elementary property  $\ve
(C X D)=(D^T \otimes C) \; \ve (X)$ in which $\otimes$ denotes the
Kronecker product (also denoted tensorial, direct or categorial
product). For a proof of this property, see Lemma 4.3.1 in
\cite{HJ2}. Applying this property to (\ref{e7}) we immediately
obtain
\begin{equation}\label{e7b}
x_{k+1}=(A\otimes B + A^T\otimes B^T) \; x_k
\end{equation}
where $x_k = \ve(X_k)$. This is the format used in the
introduction. Combining this observation with  Theorem \ref{t2} we
deduce the following theorem.\\

\begin{thm}
\label{p2}
 Let $G_A$ and $G_B$ be two graphs with adjacency
matrices $A$ and $B$, fix some initial positive matrix $Z_0>0$ and
define
$$Z_{k+1}=\frac{BZ_kA^T + B^TZ_kA}{\|BZ_kA^T + B^TZ_kA\|_F} \qquad k=0, 1, \ldots.$$
Then, the matrix subsequences $Z_{2k}$ and $Z_{2k+1}$ converge to
$Z_{even}$ and $Z_{odd}.$ Moreover, among all the matrices in the
set
$$\{ Z_{even}(Z_0), Z_{odd}(Z_0): Z_0 >0\}$$
the matrix $Z_{even}({\bf 1})$ is the unique matrix of largest
1-norm.
\end{thm}
\\

In order to be consistent with the vector norm appearing in
Theorem \ref{t2}, the matrix norm $\|.\|_F$ we use here is the
square root of the sum of all squared entries (this norm is known
as the Euclidean or Frobenius norm), and the 1-norm $\|.\|_1$ is
the sum of the magnitudes of all its entries. One can also provide
a definition of the set $Z$ in terms of one of its extremal
properties.\\

\begin{thm}
Let $G_A$ and $G_B$ be two graphs with adjacency matrices $A$ and
$B$ and consider the notation of Theorem \ref{p2}. The set
$$Z=\{ Z_{even}(Z_0), Z_{odd}(Z_0): Z_0 >0\}$$
and the set of all positive matrices that maximize the expression
$$
\frac{\|B X A^T+B^T X A\|_F}{\|X\|_F}.
$$
are equal. Moreover, among all matrices in this set there is a
unique matrix $\mathbf{S}$ whose 1-norm is maximal. This matrix
can be obtained as
$$\mathbf{S}= \lim_{k \rightarrow +\infty} Z_{2k}$$
for $Z_0={\bf 1}$.
\end{thm}

\begin{proof}  The above expression can also be written as
$\|L(X)\|_2/\|X\|_2$ which is the induced 2-norm of the linear
mapping $L$ defined by $L(X)= BX A^T + B^TXA$. It is well known
\cite{HJ} that each dominant eigenvector $X$ of ${L}^2$ is a
maximizer of this expression. It was shown above that ${\mathbf
S}$ is the unique matrix of largest 1-norm in that set.
\end{proof}
\\

We take the matrix $\mathbf{S}$ appearing in this Theorem as
definition of similarity matrix between $G_A$ and $G_B$. Notice
that it follows from this definition that the similarity matrix
between $G_B$ and $G_A$ is the transpose of the similarity matrix
between $G_A$ and $G_B$.\\

A direct algorithmic transcription of the definition leads to an
approximation algorithm for computing similarity matrices of
graphs:\\

{\tt

1. Set $Z_0 = {\bf 1}$.

2. Iterate an even number of times
$$Z_{k+1}=\frac{BZ_kA^T + B^TZ_kA}{\|BZ_kA^T + B^TZ_kA\|_F}$$
\hspace{3.2em} and stop upon convergence of $Z_k$.

4. Output $\mathbf{S}$.
}\\

This algorithm is a matrix analog to the classical power method
(see \cite{HJ}) to compute a dominant eigenvector of a matrix. The
complexity of this algorithm is easy to estimate. Let $G_A, G_B$
be two graphs with $n_A, n_B$ vertices and $e_A, e_B$ edges,
respectively. Then the products $BZ_k$ and $B^TZ_k$ require less
than $2n_A.e_B$ additions and multiplications each, while the
subsequent products $(BZ_k)A^T$ and $(B^TZ_k)A$ require less than
$2n_B.e_A$ additions and multiplications each. The sum and the
calculation of the Frobenius norm requires $2n_A.n_B$ additions
and multiplications, while the scaling requires one division and
$n_A.n_B$ multiplications. Let us define
$$ \alpha_A := e_A/n_A, \quad \alpha_B := e_B/n_B $$
as the average number of non-zero elements per row of $A$ and $B$,
respectively, then the total complexity per iteration step is of
the order of $ 4(\alpha_A+\alpha_B)n_An_B$ additions and
multiplications. As was shown in Theorem \ref{t2}, the convergence
of the even iterates of the above recurrence is linear with ratio
$(\mu / \rho )^2$. The number of floating point operations needed
to compute $\mathbf{S}$ to $\epsilon$ accuracy with the power
method is therefore of the order of
$$ 8n_An_B\frac{(\alpha_A+\alpha_B)\log \epsilon }
{(\log \mu - \log \rho)}.$$

Other sparse matrix methods could be used here, but we do not
address such algorithmic aspects in this paper. For particular
classes of adjacency matrices, one can compute the similarity
matrix $\mathbf{S}$ directly from the dominant invariant subspaces
of matrices of the size of $A$ or $B$. We provide explicit
expressions for a few such classes in the next section.

\section{Hubs, authorities and central scores}
\label{s5}

As explained in the introduction, the hub and authority scores of
a graph  can be expressed in terms of its adjacency matrix.\\

\begin{thm}
\label{t3} Let $B$ be the adjacency matrix of the graph $G_B$. The
normalized hub and authority scores of the vertices of $G_B$ are
given by the normalized dominant eigenvectors of the matrices
$BB^T$ and $B^TB$, provided the corresponding Perron root is of
multiplicity 1. Otherwise, it is the normalized projection of the
vector ${\mathbf 1}$ on the respective dominant invariant
subspaces.
\end{thm}
\\

The condition on the multiplicity of the Perron root is not
superfluous. Indeed, even for connected graphs, $BB^T$ and $B^TB$
may have multiple dominant roots: for cycle
graphs for example, both $BB^T$ and $B^TB$ are the identity matrix.\\

Another interesting  structure graph is the path graph of length
three:
\[ 1 \longrightarrow 2 \longrightarrow 3 \]

As for the hub and authority scores, we can give an explicit
expression for the similarity score with vertex $2$, a score that
we will call the {\em central score}. This central score has been
successfully used for the purpose of automatic extraction of
synonyms in a dictionary. This application is described in more
details in Section \ref{s8}.\\

\begin{thm}
\label{t4} \label{tt6} Let $B$ be the adjacency matrix of the
graph $G_B$. The normalized central scores of the vertices of
$G_B$ are given by the normalized dominant eigenvector of the
matrix $B^TB+BB^T$, provided the corresponding Perron root is of
multiplicity 1. Otherwise, it is the normalized projection of the
vector ${\mathbf 1}$ on the dominant invariant subspace.
\end{thm}
\begin{proof}
The corresponding matrix $M$ is as follows~:
\[
M=\left[ \begin{array}{ccc} 0 & B & 0 \\ B^T & 0 & B
\\ 0 & B^T & 0 \end{array} \right]
\]
and so
\[
M^2=\left[ \begin{array}{ccc} BB^T & 0 & BB \\ 0 & B^TB+BB^T & 0
\\ B^TB^T & 0 & B^TB \end{array} \right]
\]
and the result then follows from the definition of the similarity
scores, provided the central matrix $B^TB+BB^T$ has a dominant
root $\rho^2$ of $M^2$. This can be seen as follows. The matrix
$M$ can be permuted to
\[
M= P^T\left[
\begin{array}{cc} 0 & E \\ E^T & 0
\end{array}
\right]P , \quad \mbox{\rm where } E := \left[
\begin{array}{c} B
\\ B^T
\end{array}
\right]
\]

Let now $V$ and $U$ be orthonormal bases for the dominant right
and left singular subspaces of $E$ \cite{HJ}~: \begin{equation}
\label{EUV} EV = \rho U, \quad E^T U = \rho V,\end{equation} then
clearly $V$ and $U$ are also bases for the dominant invariant
subspaces of $E^TE$ and $EE^T$, respectively, since
$$
\label{MUV} E^TEV = \rho^2 V, \quad EE^T U = \rho^2 U.
$$
Moreover,
\[
P M^2 P^T = \left[
\begin{array}{cc} EE^T & 0 \\ 0 & E^T E
\end{array}
\right]
\]
and the projectors associated to the dominant eigenvalues of
$EE^T$ and $E^T E$ are respectively  $\Pi_v := VV^T$ and $\Pi_u:=
UU^T$. The projector $\Pi$ of $M^2$ is then nothing but $P^T diag
\{ \Pi_v, \Pi_u \} P$ and hence the sub-vectors of $\Pi
\mathbf{1}$ are the vectors $\Pi_v{\mathbf 1}$ and $\Pi_u{\mathbf
1}$, which can be computed from the smaller matrices $E^TE$ or
$EE^T$. Since $E^TE=B^TB+BB^T$ the central vector $\Pi_v{\mathbf
1}$ is the middle vector of $\Pi{\mathbf 1}$. It is worth pointing
out that (\ref{EUV}) also yields a relation between the two
smaller projectors~:
\[ \rho^2\Pi_v = E^T \Pi_u E, \quad \rho^2\Pi_u = E \Pi_v E^T.\]
\end{proof}
\\

\begin{figure}
\begin{center}
\includegraphics*[scale=0.5]{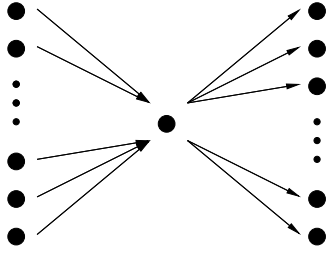} \caption{A directed bow-tie graph. Kleinberg's
hub score of the center vertex is equal to $1/\sqrt{2}$ if $m>n$
and to $0$ if $m<n$. The central score of this vertex is equal to
$1/\sqrt{2}$ independently of the relative values of $m$ and $n$.
}\label{fig:tie}
\end{center}
\end{figure}

In order to illustrate that path graphs of length 3 may have an
advantage over the \emph{hub-authority} structure graph we
consider here the special case of the ``directed bow-tie graph"
$G_B$ represented in Figure~\ref{fig:tie}. If we label the center
vertex first, then label the $m$ left vertices and finally the $n$
right vertices, the adjacency matrix for this graph is given by~
$$\label{bowty}
B =
\left[ \begin{array}{c|c|c} 0 & 0 \; \cdots \; 0 & 1 \; \cdots \; 1 \\
\hline
1 & & \\
\vdots & \mathbf{0}_n & \mathbf{0} \\
1 & & \\
\hline
0  & & \\
\vdots & \mathbf{0} & \mathbf{0}_m \\
0  & &
\end{array}\right]
$$
The matrix $B^TB+BB^T$ is equal to
$$B^TB+BB^T=\left[ \begin{array}{ccc} m+n & 0 & 0 \\
0 & \mathbf{1}_n & 0 \\ 0 & 0 &  \mathbf{1}_m \end{array}
\right],$$ and, following Theorem \ref{t4}, the Perron root of $M$
is equal to $\rho=\sqrt{n+m}$ and the similarity matrix is given
by the $(1+m+n) \times 3$ matrix $$\mathbf{S} =
\frac{1}{\sqrt{2(n+m)}}
\left[ \begin{array}{c|c|c} 0 & \sqrt{n+m} & 0  \\
\hline
1 & 0 & 0 \\
\vdots & \vdots & \vdots \\
1 & 0 & 0 \\
\hline
0 & 0 & 1 \\
\vdots & \vdots & \vdots \\
0 & 0 & 1
\end{array}\right].
$$
This result holds irrespective of the relative value of $m$ and
$n$. Let us call the three vertices of the path graph, 1, center
and 3, respectively. One could view a {\em center} as a vertex
through which much information is passed on. This similarity
matrix $\mathbf{S}$ indicates that vertex 1 of $G_B$ looks very
much like a center, the left vertices of $G_B$ look like 1's, and
the right vertices of $G_B$ look like 3's.  If on the other hand
we analyze the graph $G_B$ with the hub-authority structure graph
of Kleinberg, then the similarity scores $\mathbf{S}$ differ for
$m<n$ and $m>n$~:

$$\label{sim}
\mathbf{S}_{m>n} = \frac{1}{\sqrt{m+1}}
\left[ \begin{array}{c|c} 1 & 0 \\
\hline
0 & 0 \\
\vdots & \vdots \\
0 & 0 \\
\hline
0 & 1 \\
\vdots & \vdots \\
0 & 1
\end{array}\right], \quad
\mathbf{S}_{m<n} = \frac{1}{\sqrt{n+1}}
\left[ \begin{array}{c|c} 0 &  1  \\
\hline
1 & 0 \\
\vdots & \vdots \\
1 & 0 \\
\hline
0 & 0 \\
\vdots & \vdots \\
0 & 0
\end{array}\right].
$$

This shows a weakness of this structure graph, since the vertices
of $G_B$ that deserve the label of hub or authority completely
change between $m>n$ and $m<n$.

\section{Self-similarity matrix of a graph}
\label{s7}

\begin{figure}
{\small
\begin{tabular}{cc}
\begin{minipage}{6cm}
\begin{center}
\includegraphics[scale=0.4]{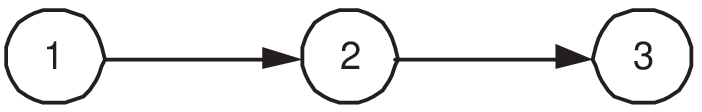}
\end{center}
\end{minipage} &
$\left[\begin{array}{ccc} 0.408&0&0\\\noalign{\medskip}0&
0.816&0\\\noalign{\medskip}0&0& 0.408\end{array}\right]$\\\\

\begin{minipage}{6cm}
\begin{center}
\includegraphics[scale=0.4]{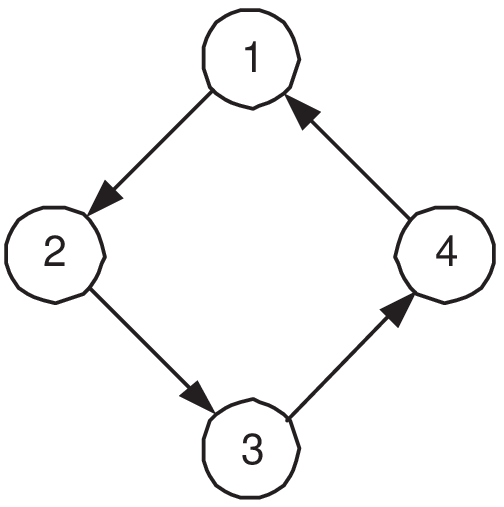}
\end{center}\end{minipage} &
$\left[\begin {array}{cccc}   0.250& 0.250& 0.250& 0.250\\
\noalign{\medskip} 0.250& 0.250& 0.250& 0.250\\\noalign{\medskip}
0.250& 0.250& 0.250& 0.250\\ \noalign{\medskip} 0.250& 0.250&
0.250& 0.250\end {array}\right]$\\\\

\begin{minipage}{6cm}
\begin{center}
\includegraphics[scale=0.4]{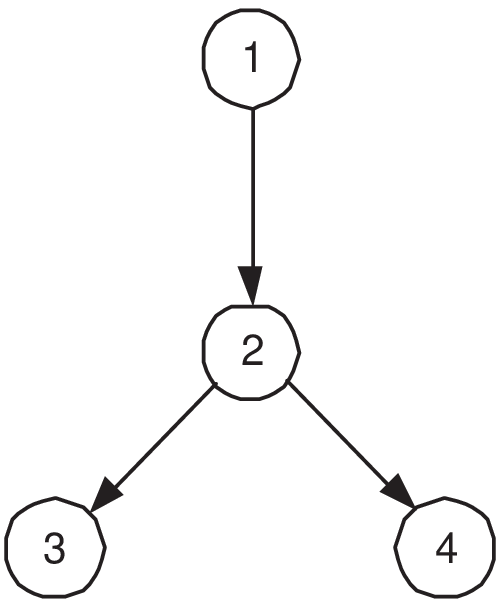}
\end{center}\end{minipage} &
$\left[\begin {array}{cccc}  0.182&0&0&0\\\noalign{\medskip}0& 0.912&0&0\\
\noalign{\medskip}0&0& 0.182& 0.182\\\noalign{\medskip}0&0& 0.182&
0.182\end {array}\right]$
\end{tabular}
}

\caption{Graphs and their corresponding self-similarity matrix.
The self-similarity matrix of a graph gives a measure of how
similar vertices are to each other.}\label{fig:sim}
\end{figure}

When we compare two equal graphs $G_A=G_B=G$, the similarity
matrix $\mathbf{S}$ is a square matrix whose entries are
similarity  scores between vertices of $G$; this matrix is the
{\it self-similarity matrix} of $G$. Various graphs and their
corresponding self-similarity matrices are represented in Figure
\ref{fig:sim}.  In general, we expect vertices to have a high
similarity score with themselves; that is, we expect the diagonal
entries of self-similarity matrices to be large. We prove in the
next theorem that the largest entry of a self-similarity matrix
always appears on the diagonal and that, except for trivial cases,
the diagonal elements of a self-similarity matrix are non-zero. As
can be seen from elementary examples, it is however not true that
diagonal elements always dominate all elements on the same row and
column.\\

\begin{thm}
The self-similarity matrix of a graph is positive semi-definite.
In particular, the largest element of the matrix appears on
diagonal, and if a diagonal entry is equal to zero the
corresponding row and column are equal to zero.
\end{thm}
\begin{proof} Since $A=B$, the iteration of the normalized
matrices $Z_k$ now becomes \[
Z_{k+1}=\frac{AZ_kA^T+A^TZ_kA}{\|AZ_kA^T+A^TZ_kA\|_F}, \quad
Z_0={\bf 1}.\] Since the scaled sum of two positive semi-definite
matrices is also positive semi-definite, it is clear that all
matrices $Z_k$ will be positive semi-definite. Moreover, positive
semi-definite matrices are a closed set and hence the limit
$\mathbf{S}$ will also be positive semi-definite. The properties
mentioned in the statement of the theorem are well known
properties of positive semi-definite matrices. \end{proof}\\

When vertices of a graph are  similar to each other, such as in
cycle graphs, we expect  to have a  self-similarity matrix with
all entries equal. This is indeed the case as will be proved in
the next section. We can also derive explicit expressions for the
self-similarity matrices of path graphs.

\begin{thm}
\label{t56}
 The self-similarity matrix of a path graph is a diagonal matrix.
\end{thm}
\begin{proof}
The product of two path graphs is a disjoint union of path graphs
and so the matrix $M$ corresponding to this graph can be permuted
to a block diagonal arrangement of Jacobi matrices
\[ J_j := \left[ \begin{array}{cccc}  0 & 1 &  \\ 1 & \ddots & \ddots \\
  & \ddots & 0 & 1 \\ & & 1 & 0\end{array}\right]\]
of dimension $j=1, \ldots \ell$, where $\ell$ is the dimension of
the given path graph. The largest of these blocks corresponds to
the Perron root $\rho$ of $M$. There is only one largest block and
its vertices correspond to the diagonal elements of $\bf S$. As
shown in \cite{Jacobi}, $\rho=2\cos(\pi/(\ell+1))$ but $M$ has
both eigenvalues $\pm \rho$ and the corresponding vectors have the
elements $(\pm)^j\sin(j\pi/(\ell+1)), \; j=1,\ldots , \ell$, from
 which $\Pi {\mathbf 1}$ can easily be computed.
\end{proof}

\section{Similarity matrices of rank one}\label{s6}

In this section we describe two classes of graphs that lead to
similarity matrices that have rank one.  We consider the case when
one of the two graphs is regular (a graph is regular if the
in-degrees of its vertices are all equal and the out-degrees are
also equal), and the case when the adjacency matrix of one of the
graphs is normal (a matrix $A$ is normal if it satisfies
$AA^T=A^TA$). In both cases we prove that the similarity matrix
has rank one. Graphs that are not directed have a symmetric
adjacency matrix and symmetric matrices are normal, therefore
graphs that are not directed always generate similarity matrices
that have rank one. \\

\begin{thm}
\label{t7} Let $G_A, G_B$ be two graphs of adjacency matrices $A$
and $B$ and assume that $G_A$ is regular. Then the similarity
matrix between $G_A$ and $G_B$ is a rank one matrix of the form
$$\mathbf{S}= \alpha \; v{\mathbf 1}^T
$$
where $v = \Pi {\mathbf 1}$ is the projection of ${\mathbf 1}$ on
the dominant invariant subspace of $(B+B^T)^2$, and $\alpha$ is a
scaling factor.
\end{thm}
\begin{proof} It is known (see, e.g.  \cite{bi}) that a regular graph
$G_A$ has an adjacency matrix $A$ with Perron root of algebraic
multiplicity 1 and that the vector $\mathbf{1}$ is the
corresponding Perron vector of both $A$ and $A^T$. It easily
follows from this that each matrix $Z_k$ of the iteration defining
the similarity matrix is of rank one and of the type $v_k {\mathbf
1}^T / \sqrt{n_A}$, where
\[ v_{k+1}=(B+B^T)v_k /\|(B+B^T)v_k\|_2, \quad v_0 = \mathbf{1}.
\] This clearly converges to $\Pi \mathbf {1}/\| \Pi
\mathbf {1}\|_2$ where $\Pi$ is the projector on the dominant
invariant subspace of $(B+B^T)^2$.
\end{proof}
\\

Cycle graphs have an adjacency matrix $A$ that satisfies $AA^T=I$.
This property corresponds to the fact that, in a cycle graph, all
forward-backward paths from a vertex return to that vertex. More
generally, we consider in the next theorem graphs that have an
adjacency matrix $A$ that is normal, i.e., that have an adjacency
matrix $A$ such that $AA^T=A^TA$.\\

\begin{thm}
Let $G_A$ and $G_B$ be two graphs of adjacency matrices $A$ and
$B$ and assume that one of the adjacency matrices is normal. Then
the similarity matrix between $G_A$ and $G_B$ has rank one.
\end{thm}
\begin{proof} Let $A$ be the normal matrix and let $\alpha$ be its Perron root.
Then there exists a unitary matrix $U$ which diagonalizes both $A$
and $A^T$~:
\[ A=U \Lambda U^*, \quad A^T = U\overline \Lambda U^*\]
and the columns $u_i, i=1,\ldots ,n_A$ of $U$ are their common
eigenvectors (notice that $u_i$ is real only if $\lambda_i$ is
real as well). Therefore
\[ (U^*\otimes I)M(U\otimes I)= (U^*\otimes I)
(A\otimes B+A^T\otimes B^T)(U\otimes I)= \Lambda \otimes B +
\overline \Lambda \otimes B^T\] and the eigenvalues of $M$ are
those of the Hermitian matrices
\[ H_i := \lambda_iB+\overline \lambda_i B^T ,\]
which obviously are bounded by $|\lambda_i|\beta$ where $\beta$ is
the Perron root of $(B+B^T)$. Moreover, if $v^{(i)}_j, j= 1,
\ldots, n_B$ are the eigenvectors of $H_i$ then those of $M$ are
given by
\[ u_i\otimes v^{(i)}_j, \quad i= 1,\ldots n_A, \quad j= 1, \ldots,
n_B \] and they can again only be real if $\lambda_i$ is real.
Since we want \emph{real} eigenvectors corresponding to extremal
eigenvalues of $M$ we only need to consider the largest real
eigenvalues of $A$, i.e. $\pm \alpha$ where $\alpha$ is the Perron
root of $A$. Since $A$ is normal we also have that its real
 eigenvectors are also eigenvectors of $A^T$. Therefore
\[ A \Pi_{\alpha} = A^T\Pi_\alpha = \alpha \Pi_\alpha , \quad
A \Pi_{-\alpha} = A^T\Pi_{-\alpha} = -\alpha \Pi_{-\alpha} .\] It
then follows that
\[ (A\otimes B+A^T\otimes B^T)^2((\Pi_{+\alpha}+\Pi_{-\alpha})\otimes \Pi_\beta) =
\alpha^2 (\Pi_{+\alpha}+\Pi_{-\alpha})\otimes \beta^2 \Pi_\beta ,
\]
and hence $\Pi :=(\Pi_{+\alpha} +\Pi_{-\alpha})\otimes \Pi_\beta$
is the projector of the dominant root $\alpha^2\beta^2$ of $M^2$.
Applying this projector to the vector $\mathbf{1}$ yields the
vector
\[ (\Pi_{+\alpha} +\Pi_{-\alpha})\mathbf{1} \otimes
\Pi_\beta\mathbf{1},\] which corresponds to the rank one matrix
\[ {\bf S} = (\Pi_{+\alpha} +\Pi_{-\alpha})\mathbf{1}
\Pi_\beta.
\]
\end{proof}

When one of the graphs $G_A$ or $G_B$ is regular or has a normal
adjacency matrix, the resulting similarity matrix ${\bf S}$ has
rank one. Adjacency matrices of regular graphs and normal matrices
have the property that the projector $\Pi$ on the invariant
subspace corresponding to the Perron root of $A$ is also the
projector on the subspace of $A^T$. As a consequence $\rho(A+A^T)
= 2 \rho(A)$. In this context we formulate the
following conjecture.\\

\begin{conj} The similarity matrix of two graphs has rank one if and only if one of the graph has the property that
its adjacency matrix $D$ is such that $\rho(D+D^T) = 2 \rho(D)$.
\end{conj}

\section{Application  to  automatic extraction of synonyms}
\label{s8}

We illustrate in this last section the use of the central
similarity score introduced in Section \ref{s5} for the automatic
extraction of synonyms from a monolingual dictionary. Our method
uses a graph constructed from the dictionary and is based on the
assumption that synonyms have many words in common in their
definitions and appear both in the definition of many words. We
briefly outline our method below and then discuss the results
obtained with the Webster dictionary on four query words. This
application given in this section is based on \cite{blond}, to
which we refer the interested reader for a complete description.

\begin{figure}
\label{lik}
\begin{center}
\includegraphics[scale=.75]{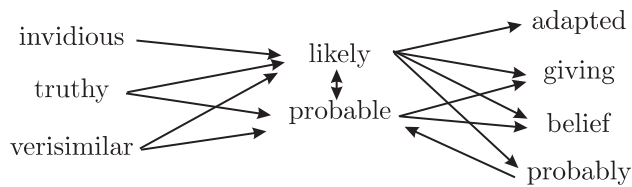}
\end{center}
\caption{Part of the neighborhood graph associated with the word
``likely". The graph contains all word used in the definition of
``likely" and all words using ``likely" in their definition.
Synonyms are identified by selecting in this graph those vertices
of highest central score.}
\end{figure}

The method is fairly simple. Starting from a dictionary, we first
construct the associated \emph{dictionary graph} $G$; each word of
the dictionary is a vertex of the graph and there is an edge from
$u$ to $v$ if $v$ appears in the definition of $u$. Then,
associated with a given query word $w$, we construct a
\emph{neighborhood graph} $G_w$ which is the subgraph of $G$ whose
vertices are pointed to by $w$ or are pointing to $w$ (see, e.g.,
Figure \ref{lik}). Finally, we compute the similarity score of the
vertices of the graph $G_w$ with the central vertex in the
structure graph
$$1\longrightarrow 2\longrightarrow 3$$ and rank the words
by decreasing score. Because of the way the neighborhood graph is
constructed, we expect the words with highest central score to be
good candidates for synonymy.

Before proceeding to the description of the results obtained, we
briefly describe the dictionary graph. We used the Online Plain
Text English Dictionary \cite{OPTED} which is based on the
``Project Gutenberg Etext of Webster's Unabridged Dictionary"
which is in turn based on the 1913 US Webster's Unabridged
Dictionary. The dictionary consists of 27 HTML files (one for each
letter of the alphabet, and one for several additions). These
files are freely available from the web site {\tt
http://www.gutenberg.net/}. The resulting graph has $112,169$
vertices and $1,398,424$ edges. It can be downloaded from the
web-page {\tt http://www.eleves.ens.fr/home/senellar/}.

In order to be able to evaluate the quality of our synonym
extraction method, we have compared the results produced with
three other lists of synonyms. Two of these (Distance and ArcRank)
were compiled automatically by two other synonym extraction
 methods (see \cite{blond} for details; the method ArcRank is
described in \cite{computer-thesaurus}), and one of them lists
synonyms obtained from the hand-made resource WordNet freely
available on the WWW, \cite{wordnet}.
  The order of appearance of the words
for this last source is arbitrary, whereas it is well defined for
the three other methods. We have not kept the query word in the
list of synonyms, since this has not much sense except for our
method, where it is interesting to note that in every example we
have experimented, the original word appears as the first word of
the list; a point that tends to give credit to our method.  We
have examined the first ten results obtained  on four query words
chosen for their variety:

\begin{enumerate}

\item \textbf{disappear}: a word with various synonyms such as
\textbf{vanish}.

\item \textbf{parallelogram}: a very specific word with no true
synonyms but with some similar words: \textbf{quadrilateral},
\textbf{square}, \textbf{rectangle}, \textbf{rhomb}\ldots

\item \textbf{sugar}: a common word with different meanings (in
chemistry, cooking, dietetics\ldots). One can expect
\textbf{glucose} as a candidate.

\item \textbf{science}: a common and vague word. It is hard to say
what to expect as synonym. Perhaps \textbf{knowledge} is the best
candidate.

\end{enumerate}

In order to have an objective evaluation of the different methods,
we have asked a sample of 21 persons to give a mark (from 0 to 10)
to the lists of synonyms, according to their relevance to
synonymy. The lists were of course presented in random order for
each word.  The results obtained are given in  the Tables
\ref{disappear}, \ref{parallelogram}, \ref{sugar} and
\ref{science}. The last two lines of each of these tables gives
the average mark and its standard deviation.

\begin{table}
\begin{center}
\footnotesize
\begin{tabular}{c|c|c|c|c}
&Distance&Our method&ArcRank&Wordnet\\
\hline
1&\textbf{vanish}&\textbf{vanish}&\textbf{epidemic}&\textbf{vanish}\\
2&\textbf{wear}&\textbf{pass}&\textbf{disappearing}&\textbf{go away}\\
3&\textbf{die}&\textbf{die}&\textbf{port}&\textbf{end}\\
4&\textbf{sail}&\textbf{wear}&\textbf{dissipate}&\textbf{finish}\\
5&\textbf{faint}&\textbf{faint}&\textbf{cease}&\textbf{terminate}\\
6&\textbf{light}&\textbf{fade}&\textbf{eat}&\textbf{cease}\\
7&\textbf{port}&\textbf{sail}&\textbf{gradually}&\textbf{}\\
8&\textbf{absorb}&\textbf{light}&\textbf{instrumental}&\textbf{}\\
9&\textbf{appear}&\textbf{dissipate}&\textbf{darkness}&\textbf{}\\
10&\textbf{cease}&\textbf{cease}&\textbf{efface}&\textbf{}\\
\hline
Mark&3.6&6.3&1.2&7.5\\
Std dev.&1.8&1.7&1.2&1.4
\end{tabular}
\caption{\label{disappear}Proposed synonyms for
\textbf{disappear}}
\end{center}
\end{table}

Concerning \textbf{disappear}, the {\em distance method} and {\em
our method} do pretty well; \textbf{vanish}, \textbf{cease},
\textbf{fade}, \textbf{die}, \textbf{pass}, \textbf{dissipate},
\textbf{faint} are very relevant (one must not forget that verbs
necessarily appear without their postposition); \textbf{dissipate}
or \textbf{faint} are relevant too. Some words like \textbf{light}
or \textbf{port} are completely irrelevant, but they appear only
in 6th, 7th or 8th position. If we compare these two methods, we
observe that our method is better: an important synonym like
\textbf{pass} gets a good ranking, whereas \textbf{port} or
\textbf{appear} are not in the top ten words. It is hard to
explain this phenomenon, but we can say that the mutually
reinforcing aspect of our method has apparently a positive effect.
In contrast to this, {\em ArcRank} gives rather poor results with
words such as \textbf{eat}, \textbf{instrumental} or
\textbf{epidemic} that are not to the point.

\begin{table}
\begin{center}
\footnotesize
\begin{tabular}{c|c|c|c|c}
&Distance&Our method&ArcRank&Wordnet\\
\hline
1&\textbf{square}&\textbf{square}&\textbf{quadrilateral}&\textbf{quadrilateral}\\
2&\textbf{parallel}&\textbf{rhomb}&\textbf{gnomon}&\textbf{quadrangle}\\
3&\textbf{rhomb}&\textbf{parallel}&\textbf{right-lined}&\textbf{tetragon}\\
4&\textbf{prism}&\textbf{figure}&\textbf{rectangle}&\textbf{}\\
5&\textbf{figure}&\textbf{prism}&\textbf{consequently}&\textbf{}\\
6&\textbf{equal}&\textbf{equal}&\textbf{parallelepiped}&\textbf{}\\
7&\textbf{quadrilateral}&\textbf{opposite}&\textbf{parallel}&\textbf{}\\
8&\textbf{opposite}&\textbf{angles}&\textbf{cylinder}&\textbf{}\\
9&\textbf{altitude}&\textbf{quadrilateral}&\textbf{popular}&\textbf{}\\
10&\textbf{parallelepiped}&\textbf{rectangle}&\textbf{prism}&\textbf{}\\
\hline
Mark&4.6&4.8&3.3&6.3\\
Std dev.&2.7&2.5&2.2&2.5
\end{tabular}
\caption{\label{parallelogram}Proposed synonyms for
\textbf{parallelogram}}
\end{center}
\end{table}

Because the neighborhood graph of \textbf{parallelogram} is rather
small (30 vertices), the first two algorithms give similar
results, which are reasonable~: \textbf{square}, \textbf{rhomb},
\textbf{quadrilateral}, \textbf{rectangle}, \textbf{figure} are
rather interesting. Other words are less relevant but still are in
the semantic domain of \textbf{parallelogram}. {\em ArcRank} which
also works on the same subgraph does not give results of the same
quality~: \textbf{consequently} and \textbf{popular} are clearly
irrelevant, but \textbf{gnomon} is an interesting addition. It is
interesting to note that {\em Wordnet} is here less rich because
it focuses on a particular aspect (\textbf{quadrilateral}).

\begin{table}
\begin{center}
\footnotesize
\begin{tabular}{c|c|c|c|c}
&Distance&Our method&ArcRank&Wordnet\\
\hline
1&\textbf{juice}&\textbf{cane}&\textbf{granulation}&\textbf{sweetening}\\
2&\textbf{starch}&\textbf{starch}&\textbf{shrub}&\textbf{sweetener}\\
2&\textbf{cane}&\textbf{sucrose}&\textbf{sucrose}&\textbf{carbohydrate}\\
4&\textbf{milk}&\textbf{milk}&\textbf{preserve}&\textbf{saccharide}\\
5&\textbf{molasses}&\textbf{sweet}&\textbf{honeyed}&\textbf{organic compound}\\
6&\textbf{sucrose}&\textbf{dextrose}&\textbf{property}&\textbf{saccarify}\\
7&\textbf{wax}&\textbf{molasses}&\textbf{sorghum}&\textbf{sweeten}\\
8&\textbf{root}&\textbf{juice}&\textbf{grocer}&\textbf{dulcify}\\
9&\textbf{crystalline}&\textbf{glucose}&\textbf{acetate}&\textbf{edulcorate}\\
10&\textbf{confection}&\textbf{lactose}&\textbf{saccharine}&\textbf{dulcorate}\\
\hline
Mark&3.9&6.3&4.3&6.2\\
Std dev.&2.0&2.4&2.3&2.9
\end{tabular}
\caption{\label{sugar}Proposed synonyms for \textbf{sugar}}
\end{center}
\end{table}

Once more, the results given by {\em ArcRank} for \textbf{sugar}
are mainly irrelevant (\textbf{property}, \textbf{grocer}, ...).
Our method is again better than the distance method:
\textbf{starch}, \textbf{sucrose}, \textbf{sweet},
\textbf{dextrose}, \textbf{glucose}, \textbf{lactose} are highly
relevant words, even if the first given near-synonym
(\textbf{cane}) is not as good. Note that our method has marks
that are even better than those of {\em Wordnet}.

\begin{table}
\begin{center}
\footnotesize
\begin{tabular}{c|c|c|c|c}
&Distance&Our method&ArcRank&Wordnet\\
\hline
1&\textbf{art}&\textbf{art}&\textbf{formulate}&\textbf{knowledge domain}\\
2&\textbf{branch}&\textbf{branch}&\textbf{arithmetic}&\textbf{knowledge base}\\
3&\textbf{nature}&\textbf{law}&\textbf{systematize}&\textbf{discipline}\\
4&\textbf{law}&\textbf{study}&\textbf{scientific}&\textbf{subject}\\
5&\textbf{knowledge}&\textbf{practice}&\textbf{knowledge}&\textbf{subject
area}\\
6&\textbf{principle}&\textbf{natural}&\textbf{geometry}&\textbf{subject
field}\\
7&\textbf{life}&\textbf{knowledge}&\textbf{philosophical}&\textbf{field}\\
8&\textbf{natural}&\textbf{learning}&\textbf{learning}&\textbf{field
of study}\\
9&\textbf{electricity}&\textbf{theory}&\textbf{expertness}&\textbf{ability}\\
10&\textbf{biology}&\textbf{principle}&\textbf{mathematics}&\textbf{power}\\
\hline
Mark&3.6&4.4&3.2&7.1\\
Std dev.&2.0&2.5&2.9&2.6
\end{tabular}
\caption{\label{science}Proposed synonyms for \textbf{science}}
\end{center}
\end{table}

The results for \textbf{science} are perhaps the most difficult to
analyze. The distance method and ours are comparable. ArcRank
gives perhaps better results than for other words but is still
poorer than the two other methods.

As a conclusion, the first two algorithms give interesting and
relevant words, whereas it is clear that ArcRank is not adapted to
the search for synonyms. The use of the central score and its
mutually reinforcing relationship demonstrates its superiority on
the basic distance method, even if the difference is not obvious
for all words. The quality of the results obtained with these
different methods is still quite different to that of  hand-made
dictionaries such as Wordnet. Still, these automatic techniques
show their interest, since they present more complete aspects of a
word than hand-made dictionaries. They can profitably be used to
broaden a topic (see the example of \textbf{parallelogram}) and to
help with the compilation of synonyms dictionaries.

\section{Concluding remarks}

In this paper, we introduce a new concept of similarity matrix and
explain how to associate a \emph{score} with the similarity of the
vertices of two graphs. We show how this score can be computed and
indicate how it extends the concept of hub and authority scores
introduced by Kleinberg. We prove several properties and
illustrate the strength and weakness of this new concept.
Investigations of properties and applications of the similarity
matrix of graphs can be pursued in several directions. We outline
some possible research directions.

One natural extension of our concept is to consider networks
rather than graphs; this amounts to consider adjacency matrices
with arbitrary real entries and not just integers. The definitions
and results presented in this paper use only the property that the
adjacency matrices involved have non-negative entries, and so all
results remain valid for networks with non-negative weights. The
extension to networks makes a sensitivity analysis possible: How
sensitive is the similarity matrix to the weights in the network?
Experiments and qualitative arguments show that, for most
networks, similarity scores are almost everywhere continuous
functions of the network entries. Perhaps this can be analyzed for
models for random graphs such as those that appear in \cite{bol}?
These questions can probably also be related to the large
literature on eigenvalues and invariant subspaces of graphs; see,
e.g., \cite{chu}, \cite{dra} and \cite{dra2}.

It appears natural to investigate the possible use of the
similarity matrix of two graphs to detect if the graphs  are
isomorphic. (The membership of the graph isomorphism problem to
the complexity classes P or  NP-complete is so far unsettled.) If
two graphs are isomorphic, then their similarity matrix can be
made symmetric by column (or row) permutation. It is easy to check
in polynomial time if such a permutation is possible and if it is
unique (when all entries of the similarity matrix are distinct, it
can only be unique). In the case where no such permutation exists
or when only one permutation is possible, one can immediately
conclude by answering negatively or by checking the proposed
permutation. In the case where many permutation render the
similarity matrix symmetric, all of them have to be checked and
this leads to a possibly exponential number of permutations to
verify. It appears interesting to see how this heuristic compares
to other heuristics for graph isomorphism and to investigate if
other features of the similarity matrix can be used to limit the
number of permutations to consider.

 More specific questions on the similarity matrix also
arise. One open problem is to characterize the pairs of matrices
that give rise to a rank one similarity matrix.  The structure of
these pairs is conjectured at the end of Section 6. Is this
conjecture correct? A long-standing graph question also arises
when trying to characterize the graphs whose similarity matrices
have only positive entries. The positive entries of the similarity
matrix between the graphs $G_A$ and $G_B$ can be obtained as
follows. First construct the product graph, symmetrize it, and
then identify in the resulting graph the connected component(s) of
largest possible Perron root. The indices of the vertices in that
graph correspond exactly to the nonzero entries in the similarity
matrix of $G_A$ and $G_B$. The entries of the similarity matrix
will thus be all positive if and only if the symmetrized product
graph is connected; that is, if and only if, the product graph of
$G_A$ and $G_B$ is weakly connected. The problem of characterizing
all pairs of graphs that have a weakly connected product was
introduced and analyzed in 1966 in \cite{har}. That reference
provides sufficient conditions for the product to be weakly
connected. Despite several subsequent contributions on this
question (see, e.g. \cite{bla}), the problem of efficiently
characterizing all pairs of graphs that have a weakly connected
product is a problem that, to our knowledge, is still open.

Another topic of interest is to investigate how the concepts
proposed here can be used, possibly in modified form, for
evaluating the similarity between two graphs, for clustering
vertices or  graphs, for pattern recognition in graphs and for
data mining purposes.

\section*{Acknowledgment}

We are pleased to acknowledge Bert Coessens of KULeuven, Belgium
for pointing out a mistake in an original version of Theorem
\ref{t56}. We also thank the reviewers for reading the manuscript
with extraordinary care and for their help in improving the
quality of the paper.

\bibliographystyle{amsplain}

\end{document}